\documentclass[runningheads]{llncs}

\usepackage{cite}
\usepackage{amsmath,amssymb,amsfonts}
\usepackage{algorithmic}
\usepackage{graphicx}
\usepackage{hyperref}
\usepackage{textcomp}
\usepackage{xcolor}
\usepackage{float}
\usepackage{caption}
\usepackage{subcaption}
\usepackage{lineno}
\usepackage{booktabs}
\usepackage{enumitem}
\usepackage{multirow}
\usepackage{hyperref}

\usepackage{array}
\newcolumntype{L}[1]{>{\raggedright\let\newline\\\arraybackslash\hspace{0pt}}m{#1}}
\newcolumntype{C}[1]{>{\centering\let\newline\\\arraybackslash\hspace{0pt}}m{#1}}
\newcolumntype{R}[1]{>{\raggedleft\let\newline\\\arraybackslash\hspace{0pt}}m{#1}}

\begin{document}

\title{First U-Net Layers Contain More Domain Specific Information Than The Last Ones}

\titlerunning{First U-Net Layers Are Domain Specific}

\author{
    Boris Shirokikh \inst{1, 2} \and
    Ivan Zakazov \inst{1, 3} \and
    Alexey Chernyavskiy \inst{3} \and
    Irina Fedulova \inst{3} \and
    Mikhail Belyaev \inst{1}
}

\authorrunning{B. Shirokikh, I. Zakazov et al}

\institute{
    Skolkovo Institute of Science and Technology, Moscow, Russia
    \and
    Kharkevich Institute for Information Transmission Problems, Moscow, Russia
    \and
    Philips Research, Moscow, Russia
    \\
    \email{Ivan.Zakazov@skoltech.ru}
}

\maketitle

\begin{abstract}
    MRI scans appearance significantly depends on scanning protocols and, consequently, the data-collection institution. These variations between clinical sites result in dramatic drops of CNN segmentation quality on unseen domains.  Many of the recently proposed MRI domain adaptation methods operate with the last CNN layers to suppress domain shift. At the same time, the core manifestation of MRI variability is a considerable diversity of image intensities. We hypothesize that these differences can be eliminated by modifying the first layers rather than the last ones. To validate this simple idea, we conducted a set of experiments with brain MRI scans from six domains. Our results demonstrate that 1) domain-shift may deteriorate the quality even for a simple brain extraction segmentation task (surface Dice Score drops from 0.85-0.89 even to 0.09); 2) fine-tuning of the first layers significantly outperforms fine-tuning of the last layers in almost all supervised domain adaptation setups. Moreover, fine-tuning of the first layers is a better strategy than fine-tuning of the whole network, if the amount of annotated data from the new domain is strictly limited. 
\end{abstract}


\section{Introduction}
\label{sec:intro}

Convolutional Neural Networks (CNN) are the most accurate segmentation methods for many medical image analysis tasks \cite{shen2017deep}. The core advantage of deep CNNs is their great flexibility due to a large number of trainable parameters. However, this flexibility may result in a dramatic drop in performance, if the test data comes from a different distribution which is a common situation for medical imaging. This fact is especially true for Magnetic Resonance Imaging (MRI) as different scanning protocols result in significant variations of slice orientation, thicknesses, and, most importantly, overall image intensities \cite{orbes2019multi,kamnitsas2017unsupervised}.  

Many of the existing MRI domain adaptation approaches rely on the information from the last CNN's layers, see details in Sec. \ref{sec:related_work}. However, we assume that the differences in intensities can be successfully reduced by modifying the first convolutional layers. Surprisingly, this simple idea has not been directly compared with other fine-tuning strategies to the best of our knowledge.
Therefore, we aimed to compare the following options in a series of supervised domain adaptation setups: fine-tuning the whole network, fine-tuning the first layers only and fine-tuning the last layers only.\\
\\
\let\labelitemi\labelitemii
\noindent Our contribution is twofold:
\begin{itemize}
    \item First, we show that publicly available dataset CC359 \cite{dataset} holds a great potential for being utilised as a benchmark dataset for testing various DA approaches. 
    \item Secondly, we prove that fine-tuning of the first layers outperforms significantly fine-tuning of the last layers. Moreover, fine-tuning of the first layers outperforms fine-tuning of the whole network when an extremely small amount of data is available.
\end{itemize}

Finally, we publish a complete experimental pipeline to provide a starting point for other researchers\footnote{https://github.com/kechua/DART20}.


\section{Related work}
\label{sec:related_work}

A lot of domain adaptation methods exploit one of the following  strategies: 
\begin{enumerate}
    \item Train a network on the source domain, then fine-tune it using data from the target domain \cite{unfrLast,unfrLast2,valverde2019one, valindria2018domain}.
   \item Remove domain-specific information using an additional network head that aims to predict the scan domain. The core idea introduced in \cite{ganin2015unsupervised} is to minimize domain prediction accuracy rather than maximize by exploiting the gradient reversal layer (GRL). Related ideas were proposed in the medical image analysis community by  \cite{kamnitsas2017unsupervised,orbes2019multi}.
  \item Various ideas around Generative Adversarial Learning, e.g., \cite{dou2018unsupervised,yang2019unsupervised}. 
\end{enumerate}
Interestingly, at least some methods from all three groups exploit information from the first/last layers explicitly or implicitly and thus are connected to the core research question of our work.

The most widespread strategy is to fine-tune the last layers of the network. Though it's a natural solution for transfer learning as the generality of features tends to decrease with the number of the layer \cite{baseUnfrLasr}, several works showed promising results for domain adaptation \cite{unfrLast,unfrLast2,valverde2019one}. In contrast, to the best of our knowledge, fine-tuning of the first layers was not properly researched for medical imaging.  Moreover, the authors of \cite{unfrLast,unfrLast2} directly rejected the idea of fine-tuning the first layers because of an assumption that they pose too general, domain-independent characteristics. GRL-related algorithms usually minimize domain shift by analyzing high-level features generated at the end of the network using similar motivation. 

In the studies \cite{lee2018simple, erdil2020unsupervised}, dedicated to out-of-distribution detection (OOD), the best performance was achieved with the confidence scores computed for the intermediate or the last layers, but in those experiments out-of-distribution and in-distribution data came from the datasets of a very different nature (e.g. CIFAR-10 and SVHN), whereas in our case an "OOD sample" would correspond to a scan from a new domain.

Meanwhile, a study \cite{unfrFirst} of non-medical images showed that domain shift effects pop up at the very first layer. The authors train the net on the base domain and then assess domain shift by observing filter maps, produced by convolutions on a particular layer. The more pronounced domain shift on a certain layer is, the more domain-specific are the distributions of the filter maps on this layer. Concluding, that the first layers are susceptible to domain shift even more than the other layers, the authors then develop an unsupervised DA method, targeting the first layer only. 

An idea somewhat similar to fine-tuning of the first layers was proposed in \cite{karani2020testtime} for the setup of test-time domain adaptation. The key element of the pipeline, which is fine-tuned during the test time, is a shallow image-to-normalized-image CNN, which may be thought of as the first layers of a net, combined from the preprocessing and the main task nets.

Finally, the authors of \cite{dou2018unsupervised} hypothesized that cross-modality (MRI-CT) domain shift causes significant changes mainly in the first layers, and developed an unsupervised domain adaptation framework based on adversarial learning. However, this idea wasn't validated directly in the paper.

\section{Experiments}
\label{sec:exp}

\subsection{Data}
\label{ssec:data}

We conduct all the experiments on a publicly available dataset CC359 \cite{dataset}. It is composed of $359$ MRIs of head acquired on scanners from three vendors (Siemens, Philips and General Electric) at both $1.5 T$ and $3 T$. Different combinations of a vendor and a field strength correspond to \textit{six domains}. Data is equally distributed across domains, with an exception of Philips $1.5 T$ domain, where only $59$ subjects are present.

We do not apply any specific preprocessing to the data, except for two simple steps. First, we transform all the scans to the equal voxel resolution of $1 \times 1 \times 1$ mm via interpolation of the slices. Secondly, we scale the resulting images to the intensities of voxels between $0$ and $1$ before passing them to the network.

\subsection{Metric}
\label{ssec:metric}


The quality of brain segmentation is usually measured with Dice Score \cite{bakas2018brats}.

\begin{equation}
\label{eq:dice_score}
\text{Dice Score} = \frac{2 \cdot |A \cap B|}{|A| + |B|}.    
\end{equation}

It measures a voxel-wise similarity between two masks $A$ and $B$:, which means that it captures volumetric quality of segmentation. However, in case of brain segmentation the most eloquent indicator of the model quality is how good the edges of the brain are segmented. Meanwhile, the edge zones account for a small share of the brain volume, which makes dice score not sensitive enough to the delineation quality. Therefore we use the Surface Dice metric \cite{sDice}, which compares how closely the segmentation and the ground truth surfaces align.

Surface Dice shares the same formula with Dice Score (Eq. \ref{eq:dice_score}), but $A$ and $B$ correspond to two surfaces in this case. The intersection of two surfaces depends on the \textit{tolerance} value, which defines the maximum distance between predicted surface voxel and the ground truth surface voxel to be considered as matching elements. We report the experimental results with the tolerance of $1$mm, since we find it sensitive enough to the changes in predictions of different methods.

\subsection{Architecture and Training}
\label{ssec:arch}

Unlike 3D CNN architectures, 2D architectures give us an opportunity to investigate the behaviour of different approaches on an extremely small amount of labeled data from the target domain, i.e. on a subset of slices from one scan (see chapter \ref{ssec:exp}), thus we use 2D U-Net \cite{ronneberger2015u} in all our experiments. Besides, we have also conducted the baseline experiments for 3D U-Net \cite{cciccek20163d} and observed equally pronounced decline in the segmentation quality.

\begin{figure}[t!]
      \includegraphics[width=\linewidth]{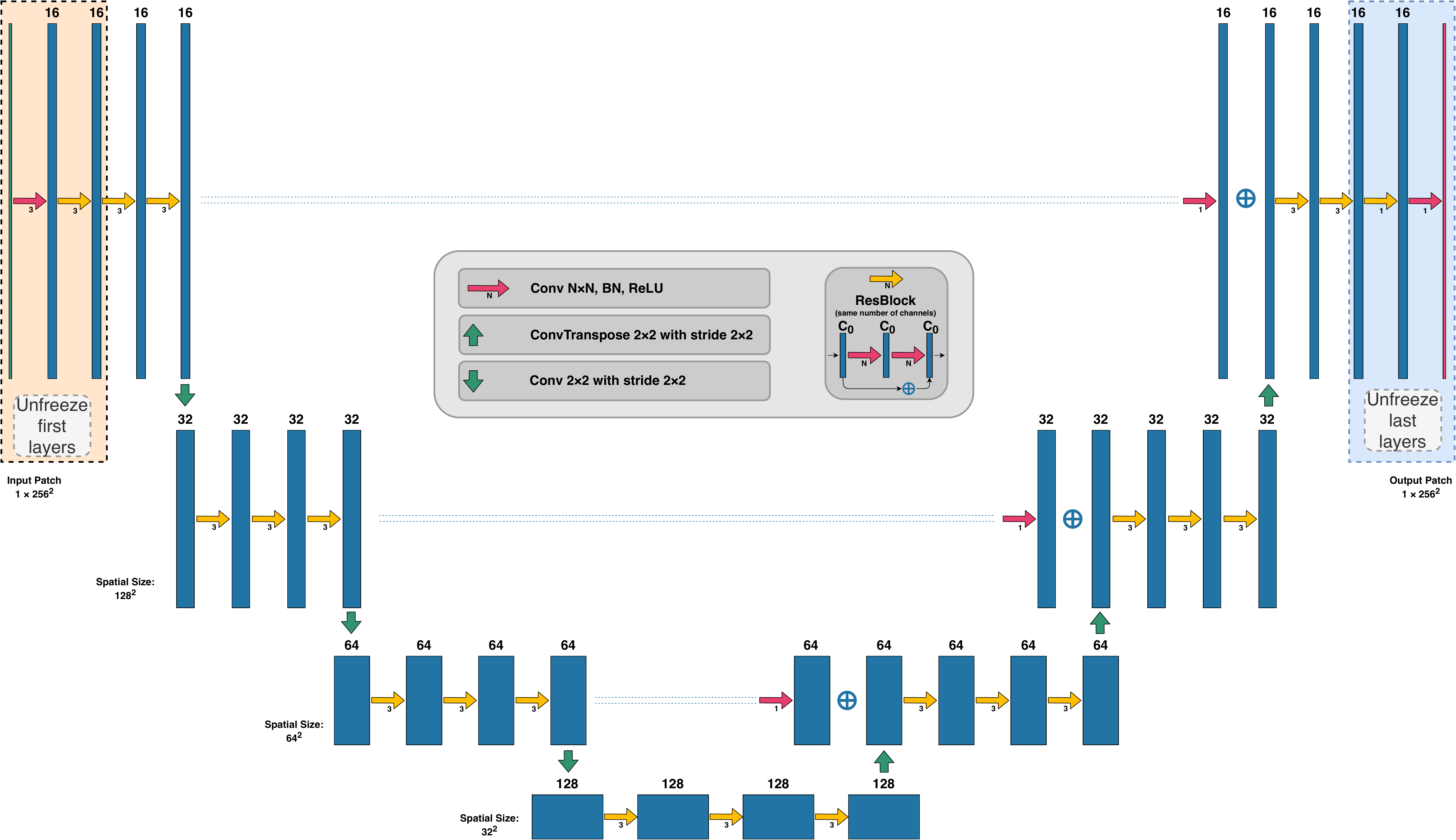}
      \caption{The architecture of 2D U-Net \cite{ronneberger2015u} with minor modification we use in our work. In the scenarios implying freezing, either the first $3$ or the last $3$ convolutional layers are unfreezed. These layers contain an equal amount of filters ($16$) of the same size, which means that in both scenarios an equal number of parameters is fine-tuned.}
      \label{fig:unet2d}
\end{figure}

U-Net \cite{ronneberger2015u} is one of the most widely used architectures which was originally developed for 2D image segmentation tasks. We adopt the original 2D U-Net model for our task introducing minor changes to keep up with the state-of-the-art level of architectures. We use residual blocks \cite{he2016deep} instead of simple convolutions, for this  was shown to improve segmentation quality \cite{milletari2016v}. We also apply convolutional layer with $1 \times 1 \times 1$ kernel to the skip-connections. We change the channel-wise concatenation at the end of the skip-connections to the channel-wise summation --  it reduces memory consumption and preserves the number of channels in the following residual blocks. Our architecture is detailed in Fig. \pageref{fig:unet2d}. We keep it without changes for the rest of the experiments. To reduce the dependence of the results on the architecture we also carried out the experiments for vanilla U-Net.

On the source domain, we train the model for $100$ epochs, starting with the learning rate of $10^{-2}$ and reducing it to $10^{-3}$ at the epoch $80$. When we transfer the model to the other domain we fine-tune it for $20$ epochs, starting with the learning rate of $10^{-3}$ and reducing it to $10^{-4}$ at the epoch $15$. Each epoch consists of $100$ iterations of stochastic gradient descent with Nesterov momentum ($0.9$). At every iteration we sample a random slice and crop it randomly to the size of $256 \times 256$. Then we form a mini-batch of size $32$ and pass it to the network.
Training for $100$ epochs takes about $4$ hours on a 16GB nVidia Tesla V100 GPU. These are the GPUs installed on the Zhores supercomputer recently launched in Skolkovo Institute of Science and Technology (Skoltech) \cite{zacharov2019zhores}.

\subsection{Experimental setup}
\label{ssec:exp}

Below we detail three main groups of the experiments we have carried out. By the term \textit{scan} we always refer to the whole 3D MRI study, while \textit{slice} is a 2D section of a \textit{scan}.

\textbf{ The baseline and the oracle.}
First of all, we have to determine whether CC359 holds a potential for being useful for DA experiments. Thus we measure cross-domain model transferability on this data set. To do so, we train separately six models within the corresponding domains, and then test each model on the other domains. This forms the \textit{baseline} of our study.
  
The test score of a model on the source domain is obtained via 3-fold cross validation. We refer to the result as \textit{oracle}. It marks the upper boundary for all transferring methods; note, though, that in some cases transferring methods may outperform the oracle.
  
In the subsequent transferring experiments the models being transferred are trained on the whole source domain. We calculate the fraction of the gap between the oracle and the baseline that the method closes, because we find this way of measuring a method performance the most interpretable (discussed in detail in Sec. \ref{sec:results}).

\textbf{Supervised DA.} In the main part of our study we consider three supervised domain adaptation strategies: fine-tuning of the whole model (\textit{all layers}), fine-tuning of the \textit{first layers} and fine-tuning of the \textit{last layers}.


Our goal is to investigate the performance of different methods under various conditions of data availability, up to extreme shortage of target domain data. Preliminary experiments shows that quality starts to deteriorate if less than 5 scans are provided, which aligns perfectly with the study \cite{valverde2019one}, where it is shown that segmentation performance decreases with the number of voxels in the ground truth mask. In case of our task, operating with a 2D network allows us to work with a subset of slices from a random scan instead of choosing a smaller ground truth mask, which is not an option.

We vary the amount of data available, starting from $3$ and $1$ additional MRI scans and then, making use of our choice of architecture, subsampling $1/2$, $1/3$, $1/6$, $1/12$, $1/24$, $1/36$ or $1/48$ axial slices from $1$ scan. In the latter scenarios we sample slices evenly with the constant step, e.g. in $1/3$ scenario we choose slices $0$, $3$, $6$ and so on.


\textbf{Discussion of the additional setups.} Despite the authors of \cite{isensee2018no} suggest focusing on the pipeline rather than the peculiarities of an architecture, we support our claim with the same line of experiments with vanilla U-Net architecture \cite{ronneberger2015u}. Moreover, extremely limited amounts of data available raise the question of augmentation, thus we repeat all the experiments for both architectures, introducing simple augmentation techniques: rotations and symmetric flips. We place all the results for vanilla U-Net and the results for the original net trained with augmentation in Supplementary Materials while discussing them in Sec. \ref{sec:results}.


In our preliminary experiments we also tried other supervised DA setups. First, instead of fine-tuning, we trained the model from scratch on joint data from the source and the target domains. Secondly, we trained the model from scratch on data from the target domain only. We do not include aforementioned strategies in the further analysis for they yield extremely poor results.
\section{Results and Discussion}
\label{sec:results}

\begin{table}[b]
\centering
\caption{Cross-domain model transfer without fine-tuning. Column names are the source domains which the model is trained on, the row name is the target domains which the model is tested on. Sm, GE, Ph correspond to vendors, i.e. Siemens, GE and Philips. Results are given in surface Dice Score and the corresponding standard deviations are placed in the brackets.}
\label{table:baseline}

\begin{tabular}{l C{1.59cm} C{1.59cm} C{1.59cm} C{1.59cm} C{1.59cm} C{1.59cm} }
\toprule
 & Sm, 1.5T & Sm, 3T & GE, 1.5T & GE, 3T & Ph, 1.5T & Ph, 3T \\
\midrule
Sm, 1.5T & \textbf{.85 (.12)} & .51 (.15) & .72 (.08) & .56 (.13) & .71 (.10) & .71 (.07) \\

Sm, 3T   & .72 (.08) & \textbf{.88 (.03)} & .70 (.07) & .67 (.10) & .63 (.10) & .66 (.06) \\

GE, 1.5T & .39 (.14) & \textit{.09 (.05)} & \textbf{.87 (.05)} & \textit{.30 (.10)} & .55 (.19) & .48 (.08) \\

GE, 3T   & .80 (.05) & .63 (.13) & .66 (.10) & \textbf{.89 (.03)} & .67 (.10) & .67 (.06) \\

Ph, 1.5T & .63 (.08) & \textit{.25 (.07)} & \textbf{.87 (.03)} & .43 (.06) & \textbf{.89 (.03)} & .46 (.08) \\

Ph, 3T   & .54 (.13) & .34 (.13) & .70 (.11) & .37 (.10) & .47 (.14) &  \textbf{.86 (.04)} \\
\bottomrule
\end{tabular}
\end{table}


We show the presence of domain shift problem in Tab. \ref{table:baseline}. Evaluating 2D U-Net model on the source domain via cross-validation yields high Surface Dice values (diagonal elements or the \textit{oracle}). Transferring the model without fine-tuning (non-diagonal elements or the \textit{baseline}) leads to considerable quality deterioration. We emphasize the best scores with bold font and the worst with italics.

To assess each DA method for a particular source-target pair we calculate the share of the \textit{gap} between the oracle and the baseline this method closes. We denote this share $D_R$ and define it the following way:

\begin{equation}
D_R = \dfrac{D - D_B}{D_O - D_B} = \dfrac{\Delta(transferring)}{\Delta(oracle)},
\end{equation}

where $D_O$ is the oracle Surface Dice score on the target domain, $D_B$ is the baseline score on the target domain and $D$ is the score of a method being considered. 

Below we compare three chosen approaches to supervised DA problem:
fine-tuning of the whole model (\textit{all layers}), fine-tuning of the \textit{first layers} and fine-tuning of the \textit{last layers}. 


We consider the dependence of the relative improvement score on the amount of target training data. We average the scores across $30$ possible pairs of source-target domains, excluding the same-domain inference and depict the trend in Fig. \ref{fig:gap}. We also depict the density distributions of the scores.

\begin{figure}[t!]
      \includegraphics[width=\linewidth]{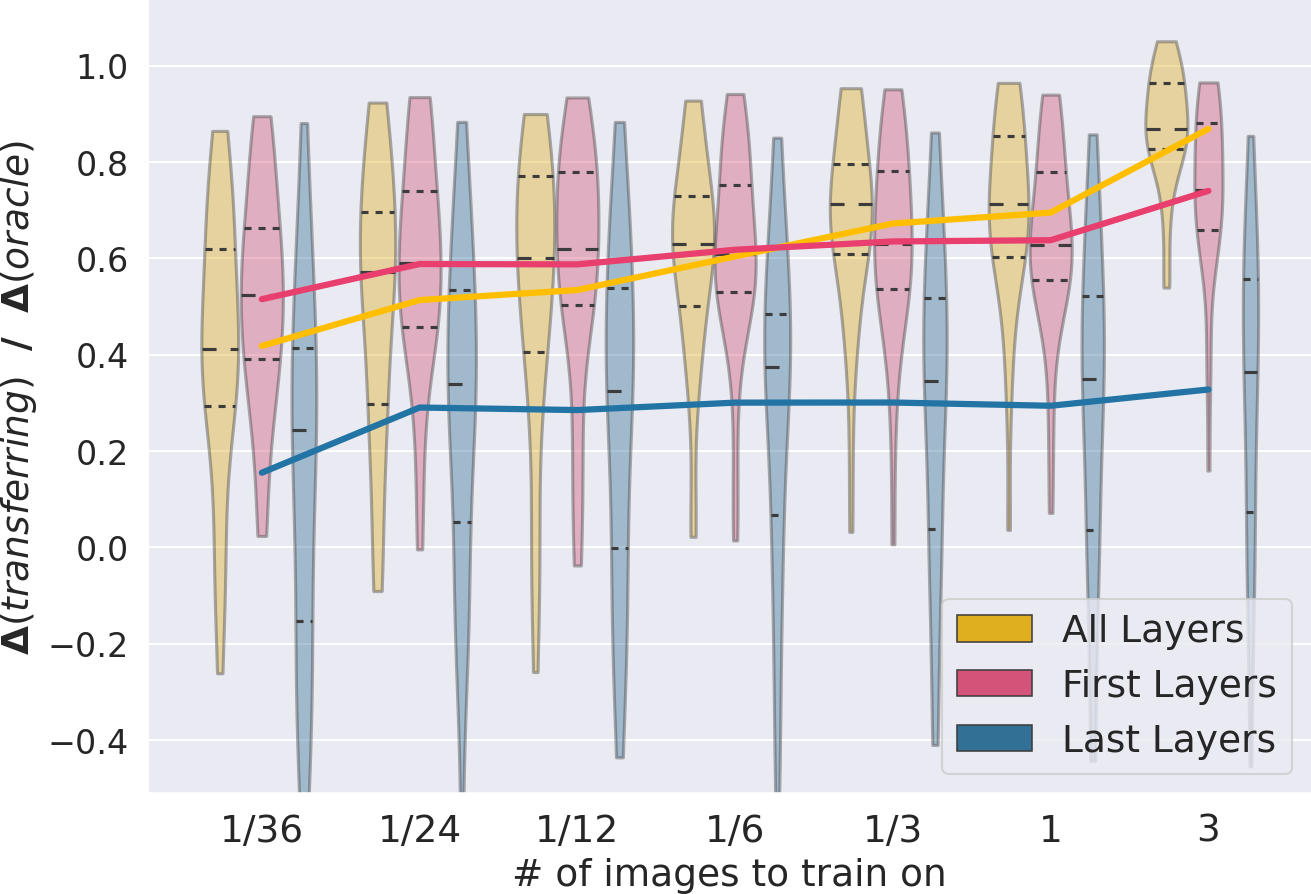}
      \caption{Dependence of the relative Surface Dice improvement (y-axis) on the target domain data availability (x-axis) for the three transferring strategies. The lines correspond to the average scores. We also include distribution densities on $30$ source-target domain pairs for every strategy.}
      \label{fig:gap}
\end{figure}

In Fig. \ref{fig:winners} we report the number of the source-target pairs on which a selected method outperforms all the other methods (sums up to $30$ over all methods for each set-up). We use paired sign test for every source-target pair to calculate the significance level. The instances for the sign test are the relative improvements of the Surface Dice scores different DA approaches yield for every single test image of the target domain.

Contrary to a mainstream conception, we show that fine-tuning of the first layers outperforms considerably fine-tuning of the last layers in our task. We therefore argue, that low-level features corresponding to the image intensity profile could be re-learned more efficiently than high-level features, which correspond to different brain structures and distinctive shapes.



Aside from freezing strategies comparison, we may see that under scarce data condition fine-tuning of the first layers becomes superior to fine-tuning of the whole model. It makes the former approach preferable in a highly practical setup, corresponding to the lack of annotated data in the target domain.

Substituting U-Net with residual blocks with vanilla U-Net or adding augmentation to either of them does not change the trends described. The results for those setups may be found in Supplementary materials.

\begin{figure}[H]
      \includegraphics[width=\linewidth]{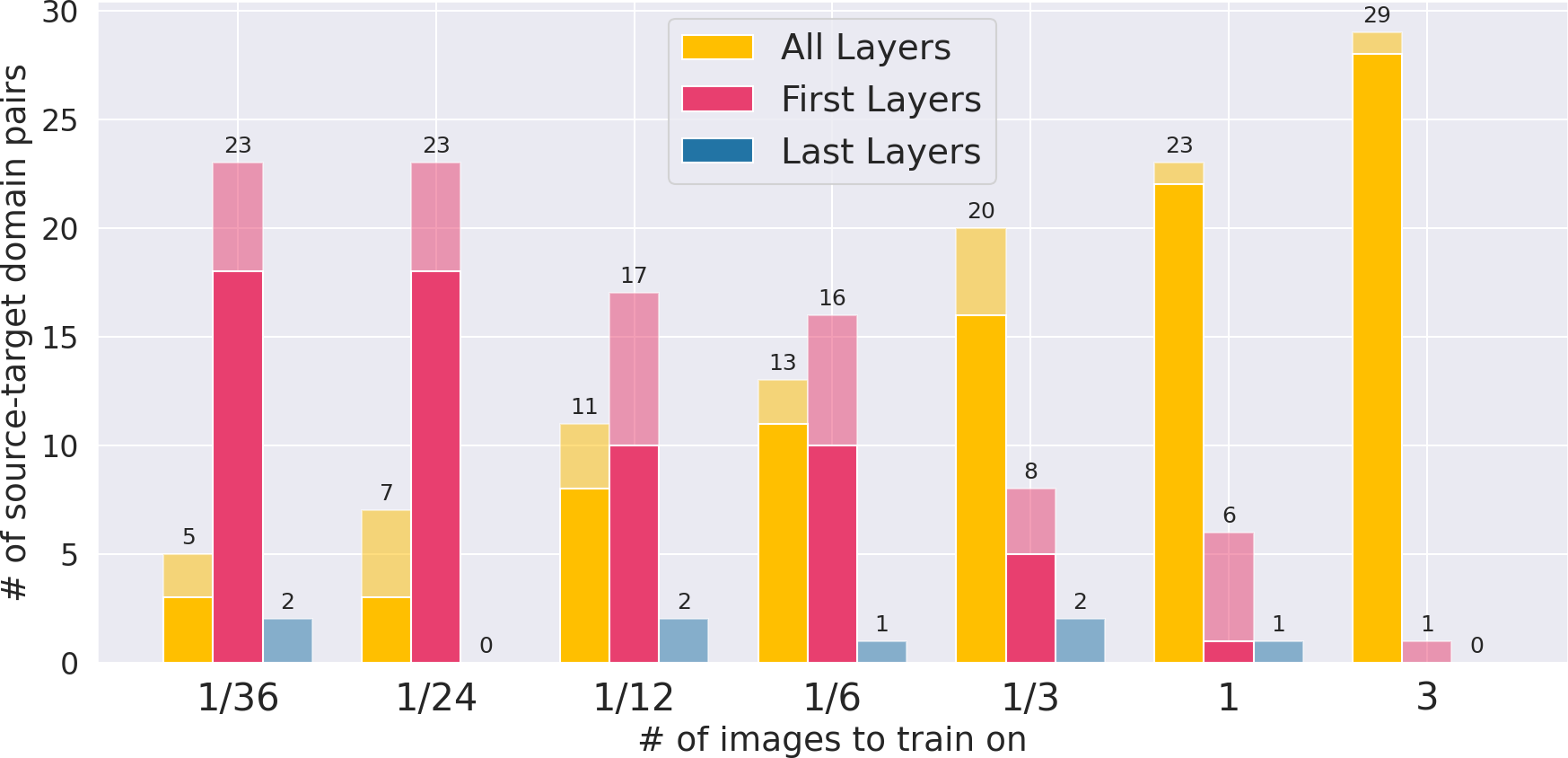}
      \caption{Dependence of the best domain adaptation strategy on the target domain data availability. Each bar counts the number of target domains which the corresponding method outperforms on. The shadowed parts of the bars correspond to the cases, when the approach outperforms the others by the average score but below the statistical significance level (p-value of paired Sign Test $< 0.1$).}
      \label{fig:winners}
\end{figure}

\section{Conclusion}
\label{sec:conclusion}

We show a drastic reduction in segmentation quality for a naive model transfer between the domains of $CC359$. We hypothesize that the low-level feature maps of this data set are more prone to domain shift than the feature maps of deeper layers; hence the first layers are the primary source of the performance degradation. 

We show that to be true by comparing different approaches to fine-tuning the network: fine-tuning the first layers outperforms fine-tuning the last layers. We also find that under the lack of annotated data for the target domain, fine-tuning of the first layers is superior to fine-tuning of the whole network.

Though we investigate a simple supervised setup, our results may suggest that unsupervised approaches will also benefit from targeting the first layers rather than the last ones.

\bibliographystyle{splncs04}
\bibliography{main.bib}

\newpage

\textbf{\LARGE Supplementary Materials}
\vspace{10mm}

In this section we present the results of the additional experiments we have carried out in order to substantiate the claims of the paper. They fall into two categories:

\begin{itemize}
  \item The same experiments as described in the paper conducted with Vanilla 2D U-Net
  \item Experiments with either an original net or Vanilla U-Net conducted with augmentation
\end{itemize}

\textbf{1.  Vanilla 2D U-Net Experiments}
\label{sec:supplementary}

\begin{figure}[h]
      \includegraphics[width=\linewidth]{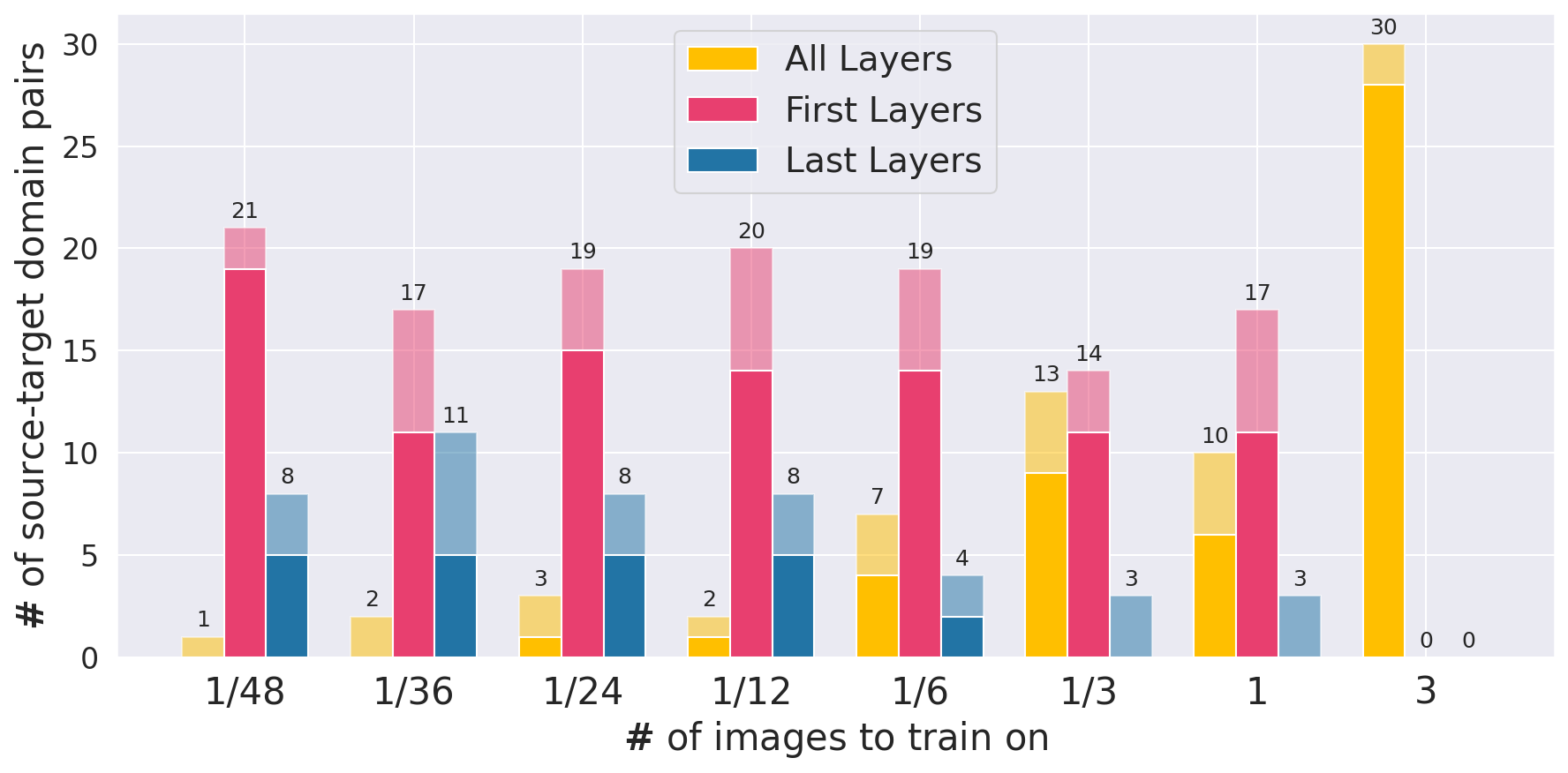}
      \caption{Dependence of the best domain adaptation strategy on the target domain data availability. Each bar counts the number of target domains which the corresponding method outperforms on. The shadowed parts of the bars correspond to the cases, when the approach outperforms the others by the average score but below the statistical significance level (p-value of paired Sign Test $< 0.1$).}
      \label{fig:winners}
\end{figure}

\begin{figure}[H]
      \includegraphics[width=\linewidth]{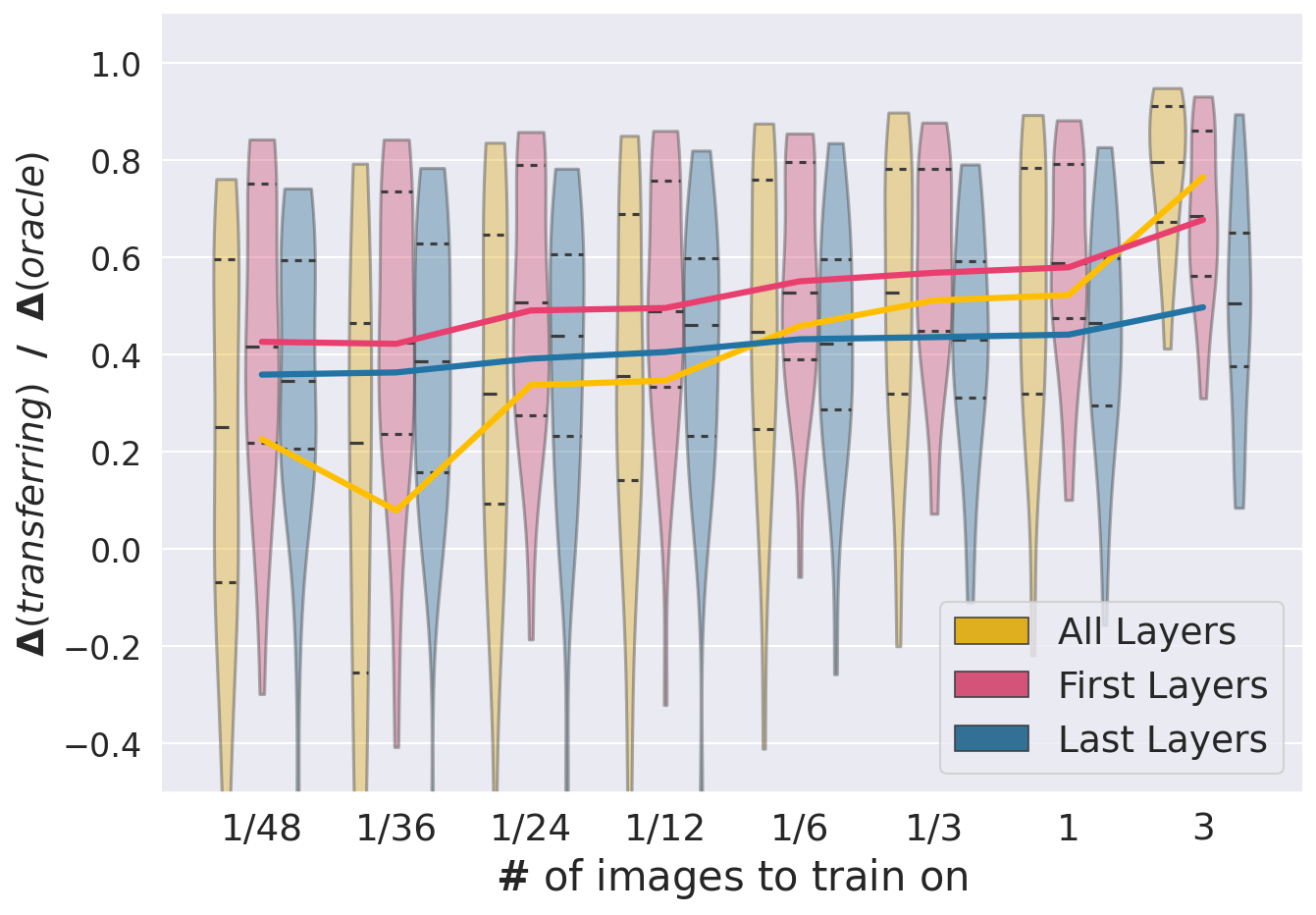}
      \caption{Dependence of the relative Surface Dice improvement (y-axis) on the target domain data availability (x-axis) for the three transferring strategies. The lines correspond to the average scores. We also include distribution densities on $30$ source-target domain pairs for every strategy.}
      \label{fig:gap}
\end{figure}






\newpage
\textbf{2. Augmentation Experiments}

\begin{figure}[h]
     \begin{subfigure}{\textwidth}
      \includegraphics[width=\linewidth]{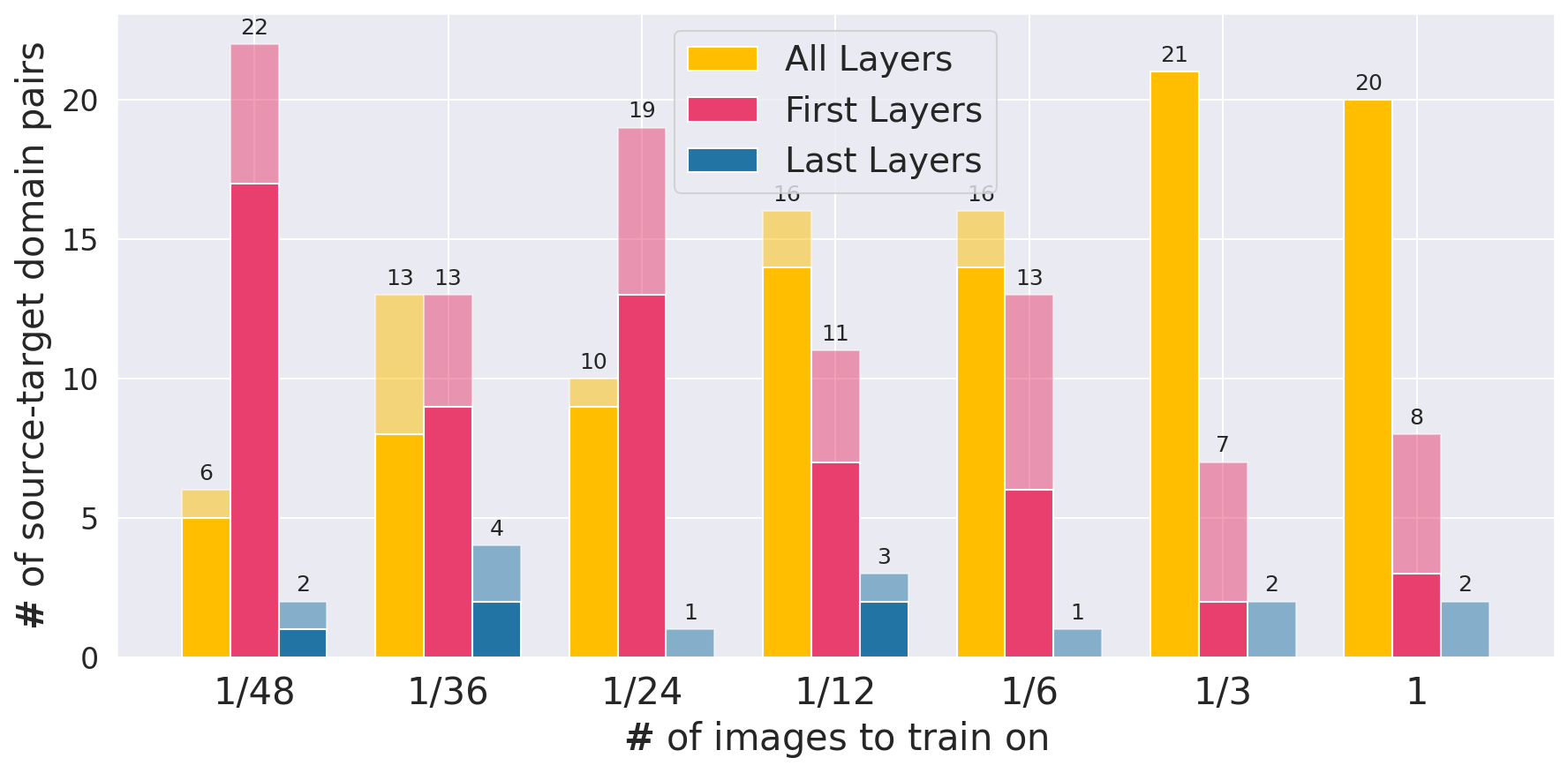}
    \end{subfigure}

    \begin{subfigure}{\textwidth}
      \includegraphics[width=\linewidth]{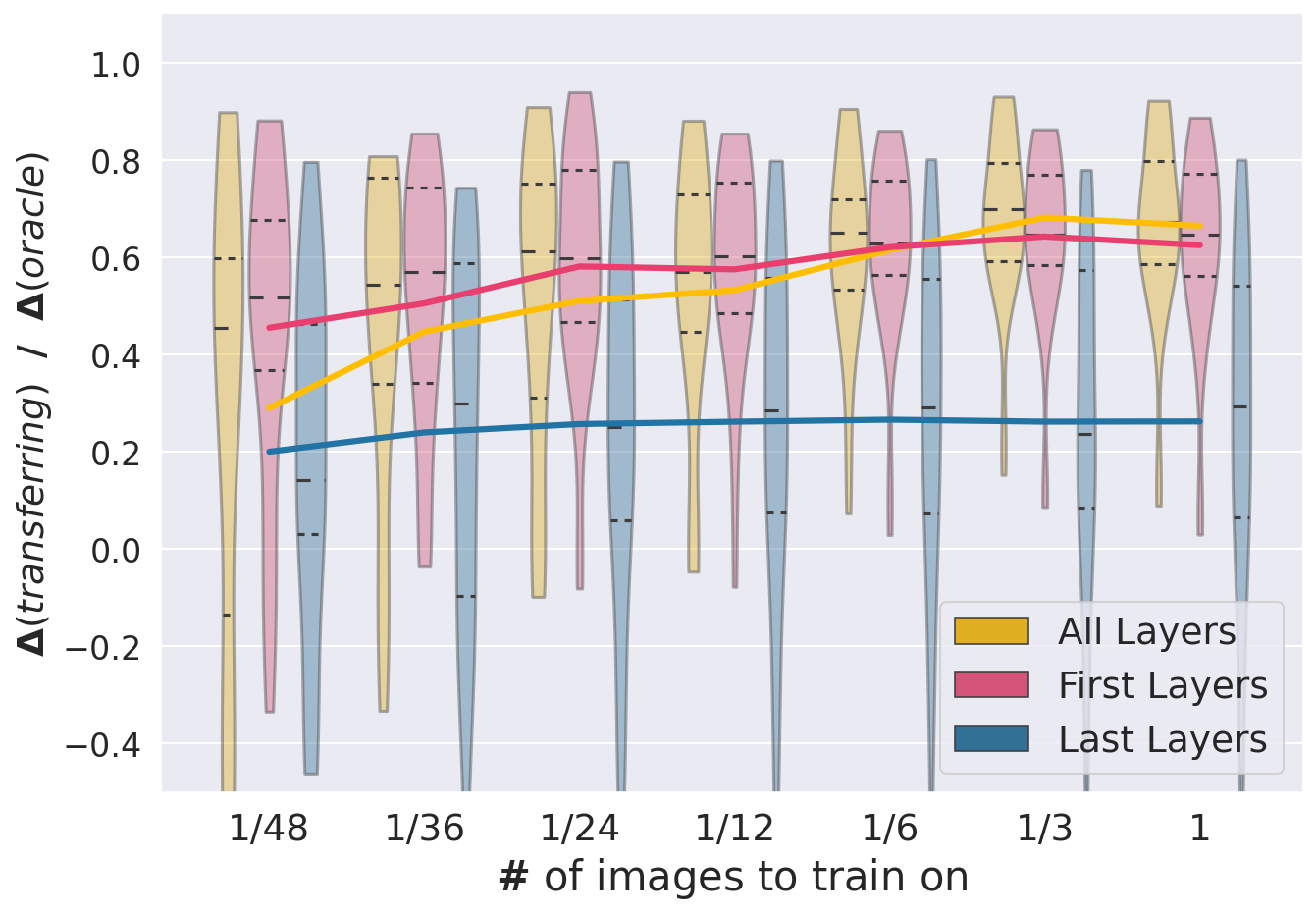}
    \end{subfigure}
    
      \caption{Results for 2D U-Net with residual blocks trained with augmentation}
      \label{fig:gap}
      
\end{figure}

\begin{figure}[H]
     \begin{subfigure}{\textwidth}
      \includegraphics[width=\linewidth]{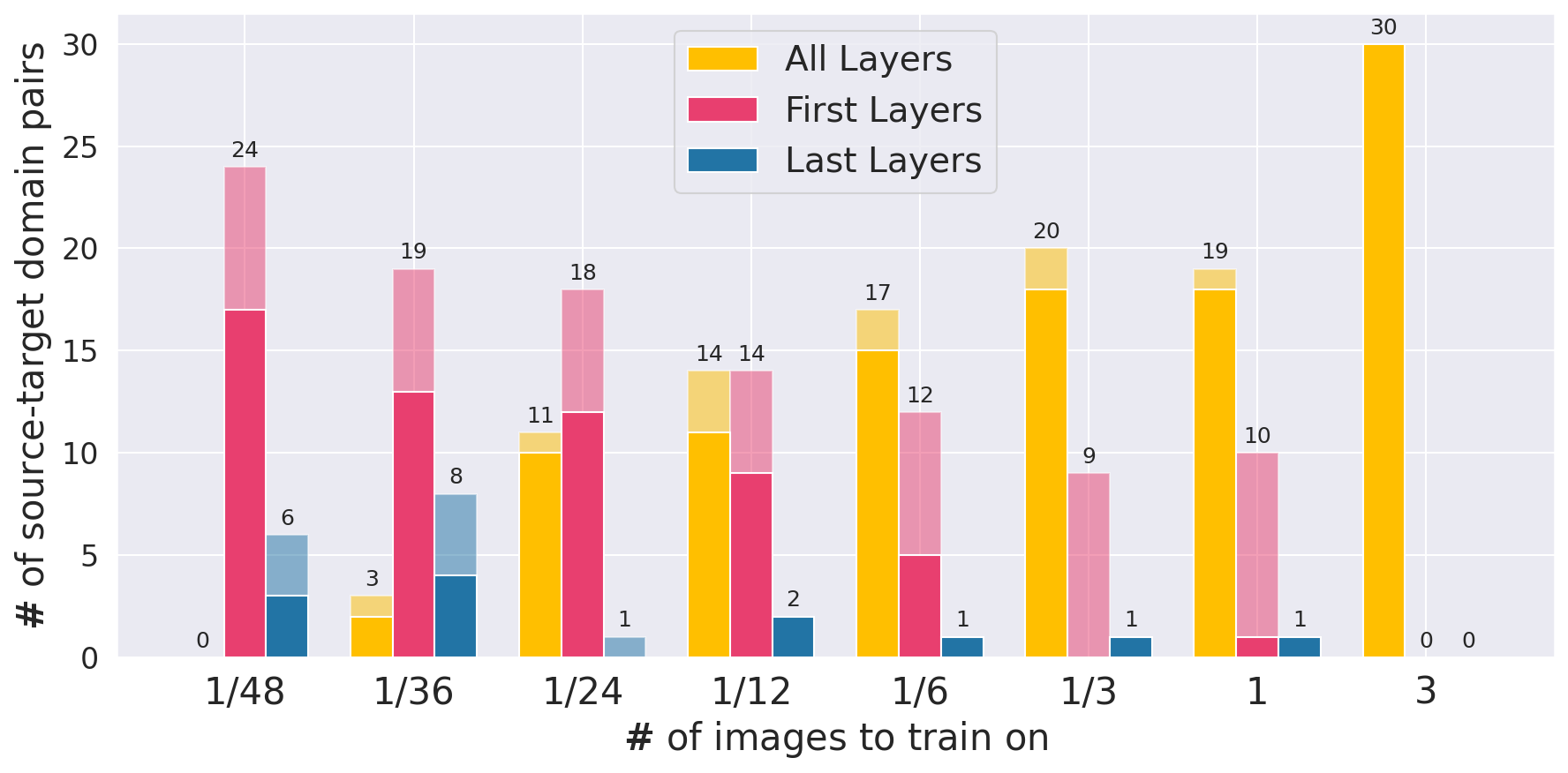}
    \end{subfigure}

    \begin{subfigure}{\textwidth}
      \includegraphics[width=\linewidth]{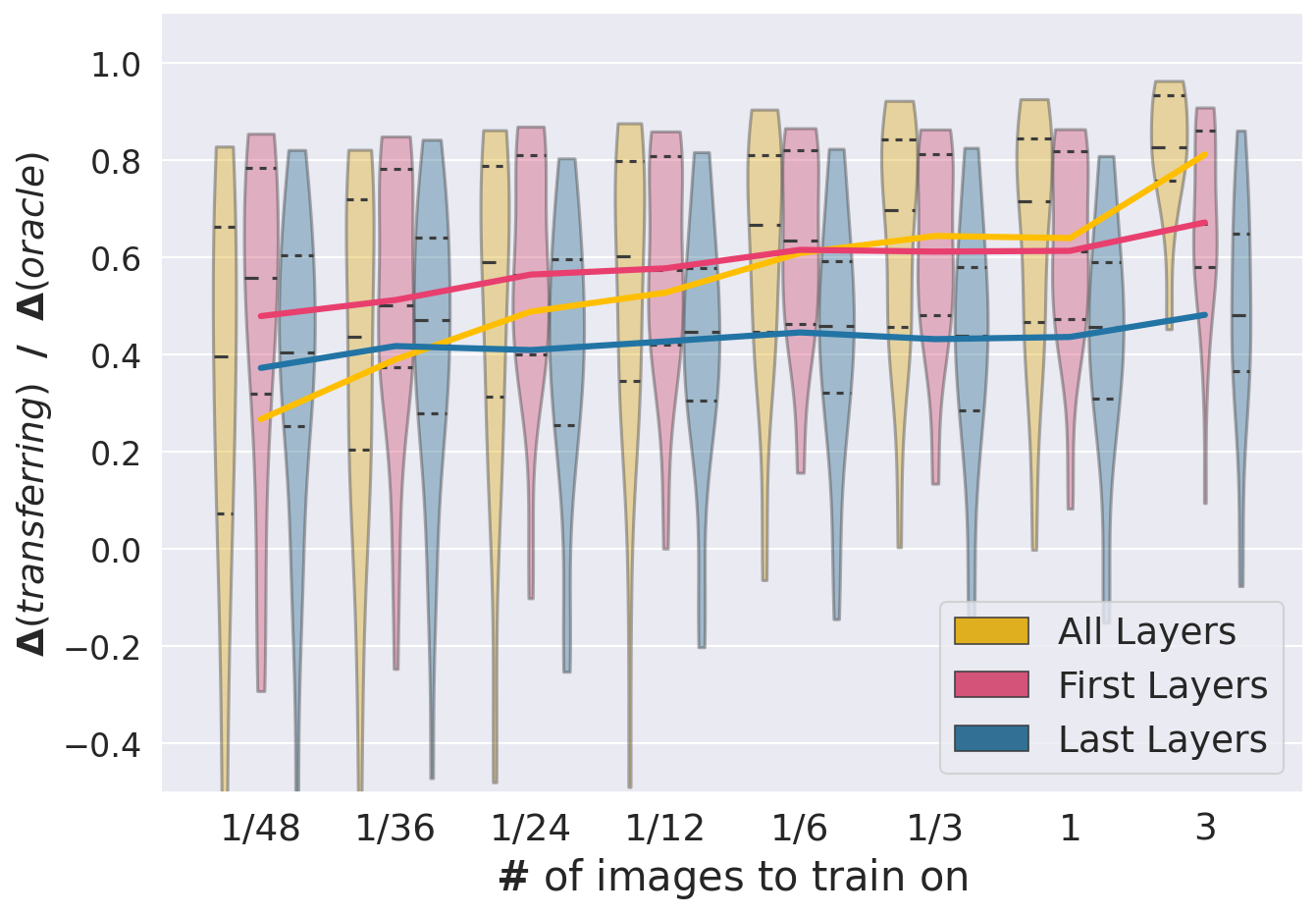}
    \end{subfigure}
    
      \caption{Results for 2D vanilla U-Net trained with augmentation}
      \label{fig:gap}
      
\end{figure}

\end{document}